\documentclass[letter]{elsart}
\usepackage{graphicx}
\usepackage{amssymb}
\begin{document}
\begin{frontmatter}
\title{\emph{Ab initio} molecular dynamics study of liquid methanol}
\author{Jan-Willem Handgraaf, Titus S. van Erp, and Evert Jan Meijer}
\address{ Department of Chemical Engineering, University of Amsterdam,
 Nieuwe Achtergracht 166, 1018 WV AMSTERDAM, The Netherlands\\
 Fax: +31-20-5255604, email: ejmeijer@science.uva.nl} 
\begin{abstract}
  We present a density-functional theory based molecular-dynamics
  study of the structural, dynamical, and electronic properties of
  liquid methanol under ambient conditions. The calculated radial
  distribution functions involving the oxygen and hydroxyl hydrogen
  show a pronounced hydrogen bonding and compare well with recent
  neutron diffraction data, except for an underestimate of the
  oxygen-oxygen correlation.  We observe that, in line with infrared
  spectroscopic data, the hydroxyl stretching mode is significantly
  red-shifted in the liquid. A substantial enhancement of the dipole
  moment is accompanied by significant fluctuations due to thermal
  motion. Our results provide valuable data for improvement of empirical
  potentials.
\end{abstract}
\maketitle
\end{frontmatter}

\section*{\normalsize\bf 1. Introduction}
Liquid methanol is of fundamental interest in natural sciences and
of significant importance in technical and industrial
applications. The liquid phase of the simplest alcohol is widely
studied, both experimentally and theoretically.
Among the alcohols, methanol is the closest analog to water.  The
characteristic hydroxyl group allows methanol to form hydrogen bonds
that dominate the structural and dynamical behavior of the liquid
phase.  The methyl group does not participate in the hydrogen bonding
and constitutes the distinction with water.  This difference is
apparent in the microscopic structure of the liquid, with water having
a tetrahedral-like coordination, whereas for methanol experiments and
molecular simulation suggest a local structure consisting of chains,
rings, or small clusters.

The precise quantification of the microscopic structural and dynamical
picture of liquid methanol has been a long-time subject in both
experimental and molecular simulation studies.  Recently, a series of
state-of-the-art studies have been reported. Among these are the
neutron diffraction (ND) experiments of Refs.~\cite{YaHi99,AdBi00}.
Simulation studies include work based on empirical
force-fields\cite{HaFe87,BiKa00}, mixed empirical and ab-initio
interactions\cite{TuLa01,MaSa02}, and a full ab initio molecular
dynamics study\cite{TsKa99}.  The recent ND experiments have provided
a detailed microscopic picture of the structure of liquid methanol,
including the pair distribution functions among all atoms.  Yet,
some of these atom-atom distribution functions are still subject to some
uncertainty as they are obtained indirectly.

Molecular simulation provides a complementary
approach to study the microscopic behavior of liquids. Most molecular
simulations studies of liquid methanol are based on empirical
force fields potentials that are designed to reproduce a selection
of experimental data.  Obviously, molecular simulations based on these
potentials do not provide a picture completely independent from
experiment. Moreover, the reliability of the results at conditions
that are significantly different from those where the potential was
designed for, may be questionable.  Density functional theory (DFT)
based molecular dynamics (MD) simulation, such as the Car-Parrinello
molecular dynamics method\cite{CaPa85}, where the interactions
are calculated by accurate electronic structure calculations provides
a route to overcome these limitations. This has been demonstrated in
studies of liquid water\cite{LaSp93,SpHu96,SiPa99-1} and aqueous
solvation\cite{MaSp97,ErMe01,RaKl02}.  Important advantages of DFT-MD
over force-field MD are that it intrinsically incorporates
polarization, that it accounts for the intra-molecular motion and
therefore allows for a direct comparison with spectroscopy of
intra-molecular vibrations, and that it yields detailed information on
the electronic properties, such as the energy levels of electronic states and
the charge distribution. In a broader chemical perspective it is
important to note that DFT-MD is capable of the study of chemical
reactions in solution, where force-field MD would fail completely as
it cannot account for the change in chemical bonding.

Here, we report a DFT-MD study of liquid methanol that addresses the
liquid structure, the inter- and intra-molecular dynamics, and the
electronic charge distribution.

\section*{\normalsize\bf 2. Methods and Validation}
Electronic structure calculations are performed using the Kohn-Sham
formulation of DFT. We employed the gradient-corrected BLYP
functional\cite{LeYa88,Beck88_2}.
The choice for the BLYP functional was guided by
its good description of the structure and dynamics of
water\cite{SpHu96} where hydrogen bonds are, as in liquid
methanol, the dominant interactions. Furthermore, it has been shown
that DFT-BLYP gives a proper description of solvation of methanol in
water\cite{ErMe01}.

The DFT-based MD simulations are performed with the Car-Parrinello
method \cite{CaPa85,MaHu00} using the CPMD package\cite{CPMD33}. Semi-local
norm-conserving Martins-Troullier pseudopotentials\cite{TrMa91} are
used to restrict the number of electronic states to those of the
valence electrons.  The pseudopotential cut-off radii were taken
$0.50$, $1.11$ and $1.23$~a.u, for H, O, and C, respectively. The
electronic states are expanded in a plane-wave basis with a cut-off of
$70$~Ry yielding energies and geometries converged within
$0.01$~\AA~and $1$~kJ/mol, respectively.  Vibrational frequencies are
converged within $1$~\%, except for C-O and O-H stretch modes that are
underestimated by $3$~\% and $5$~\% compared to the basis-set limit
values\cite{ErMe01}.

\begin{figure}[ht!]
\includegraphics[angle=0,width=14.5cm]{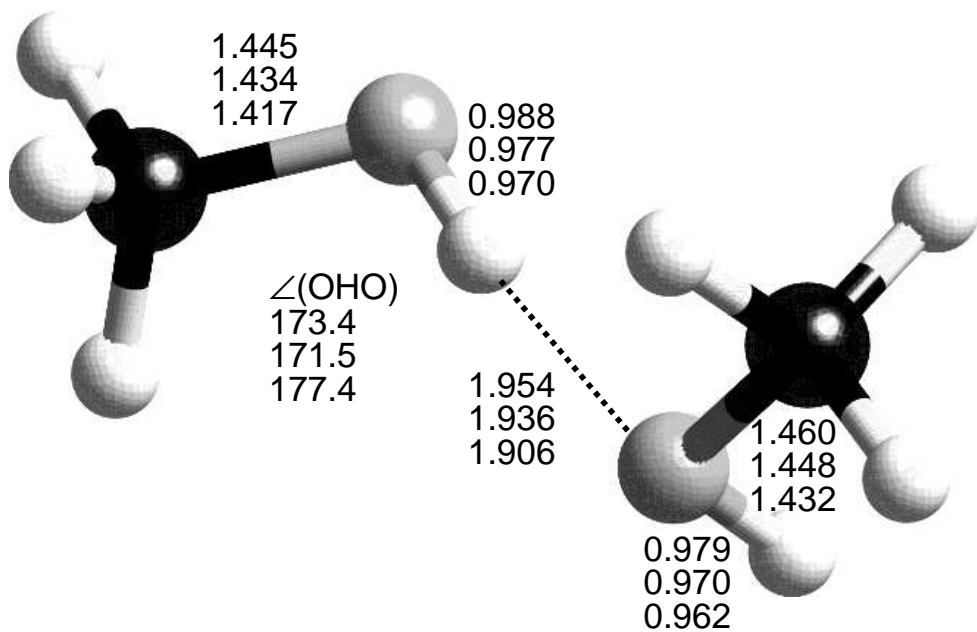}
\caption{Optimized geometry of the methanol dimer. Selected distances (\AA) and angles
  (degrees) are given for three computational methods: CPMD-BLYP (top,
  this work), ADF-BLYP\cite{ADF2000} (second, this work) and
  B3LYP\cite{MoYa97} (third).  The MP2 results of Ref.~\cite{MoYa97}
  are within 0.01~\AA~of the B3LYP result.}
\label{fig:dimer}
\end{figure}

To validate our computational approach we compared results for the
gas-phase monomer and hydrogen-bonded dimer against state-of-the-art
atomic-orbital DFT calculations obtained with
ADF\cite{ADF2000},\footnote{
Kohn-Sham orbitals are expanded in an even-tempered, all-electron Slater type
basis set augmented with 2p and 3d polarization functions for H and 3d and 4f
polarization functions for C and O.}
and against the B3LYP and MP2 calculations of Ref.~\cite{MoYa97}.
The CPMD-BLYP calculations were
performed using a cubic box with an edge of 12.9~\AA, with the
interactions among the periodic images eliminated by a screening
technique\cite{MaHu00}.  Results for geometry and complexation energy
of the dimer are given in Fig.~\ref{fig:dimer} and
Table~\ref{tab:dimer}.  Deviations among CPMD and ADF are 1~kJ/mol for
the complexation energy, smaller than 0.01~\AA\ for the
intra-molecular bonds, and smaller than 0.02~\AA\ for the hydrogen
bond.  This indicates a state-of-the-art accuracy for the numerical
methods employed in CPMD.  Compared to the MP2 and B3LYP results, the
BLYP bond lengths are slightly longer, with deviations up to $0.03~$
and $0.05$~\AA\ for the intra- and inter-molecular bond lengths,
respectively.  Differences among BLYP, B3LYP and MP2 complexation
energies are within acceptable limits, with the BLYP energies smaller
by 2-4~kJ/mole.  The deviations are similar to the comparison between
BLYP\cite{SpHu96} and MP2\footnote{MP2 limit estimate. See for example
  \cite{SchBr97}} for the water dimer and the water-methanol
dimer\cite{GoMo98,ErMe01}.  We obtained a zero-Kelvin association
enthalpy $\Delta H^0(0)$ of $10.6$~kJ/mol using the B3LYP zero-point
energy of Ref.~\cite{MoYa97}.  This is in reasonable agreement with
the experimental value of 13.2(4) kJ/mol.  The calculated hydrogen
bond length ($r_{\mathrm{O\ldots O}}=2.94$~\AA, $r_{\mathrm{H\ldots
    O}}=1.95$~\AA) is in good agreement with the experimental values
of $r_\mathrm{O\ldots O}=2.98(2)$\AA\ and $r_\mathrm{H\ldots
  O}=1.96(2)$~\AA\ of Refs.~\cite{LoBe95} and \cite{LoHa97},
respectively. 

Current gradient corrected functionals such as BLYP do not account for
dispersion forces. For methanol this could be important as attraction
to the methyl group is fully due to the dispersion force.  To estimate
the effect of the absence of the dispersion we computed the BLYP
binding energy of two dimer configurations that are sensitive to this:
one with methyl groups approaching (M-M) and the other with the
methyl and hydroxyl group approaching each other (M-OH).  State-of-the
art MP2 calculations\cite{MoDu99}, that incorporate to a good
approximation the dispersion force, serves as a reference. The dimer
configurations were taken from Ref.~\cite{MoDu99} and chosen such that
the carbon-carbon (M-M dimer) and carbon-oxygen (M-OH dimer) distances
were close to the peak position of their atom-atom distribution
function in the liquid state.\footnote{
Configuration from Ref.~\cite{MoDu99}. M-M dimer: geometry M with
$r_{\mathrm{CC}}=3.75$~\AA. M-OH dimer: geometry I with
$r_{\mathrm{CO}}=3.50$~\AA. Both monomers in the dimer are kept fixed to
their isolated geometries.} The comparison
yields for the M-M dimer values of $2.3$~kJ/mol and $-2.0$~kJ/mol for BLYP
and MP2, respectively.  For the M-OH dimer these values are
$-1.2$~kJ/mol and $-4.9$~kJ/mol, respectively. The too repulsive nature of
the BLYP interaction is consistent with DFT calculations of dispersion
dominated systems\cite{KrPu94,MeSp96}.  However, although by far not
insignificant, the magnitude of the deviation is much smaller than the
hydrogen-bond interaction and of the same order of magnitude as the
error in the latter. It can therefore be argued that for a study of
liquid methanol on the accuracy level of BLYP, neglecting the dispersion
interaction is acceptable.

In Ref.~\cite{ErMe01} we have shown that for the
gas-phase monomer CPMD-BLYP vibrational frequencies are in excellent
agreement with ADF results and, compared to experiment, underestimate
allmost all modes by $\approx~10\%$, a known feature of the BLYP
functional.

Overall, we conclude that our level theory is satisfactory in
comparison with experimental and other theoretical gas-phase data.

\section*{\bf\normalsize 3. Liquid}
Liquid methanol was modeled by 32 molecules in a periodic
cubic box with an edge of 12.9 \AA, reproducing the experimental density of
0.791~g/cm$^3$ at 293~K\cite{RiBu86}. 
The temperature was fixed at 293 K using the Nos\'e-Hoover thermostat
\cite{Nose84_1}.  The fictitious mass associated with the plane-wave
coefficients is chosen at 900 a.u., which allowed for a time step in
the numerical integration of the equations-of-motion of 0.145~fs.  The system
was equilibrated for 1 ps from an initial configuration
obtained from a force-field simulation.  Subsequently, we gathered
statistical averages from a 6 ps trajectory.

\subsection*{\rm\em\normalsize 3.1 Structure}\label{sec:Results-A}

\begin{figure}[ht!]
\includegraphics[angle=0,width=15cm]{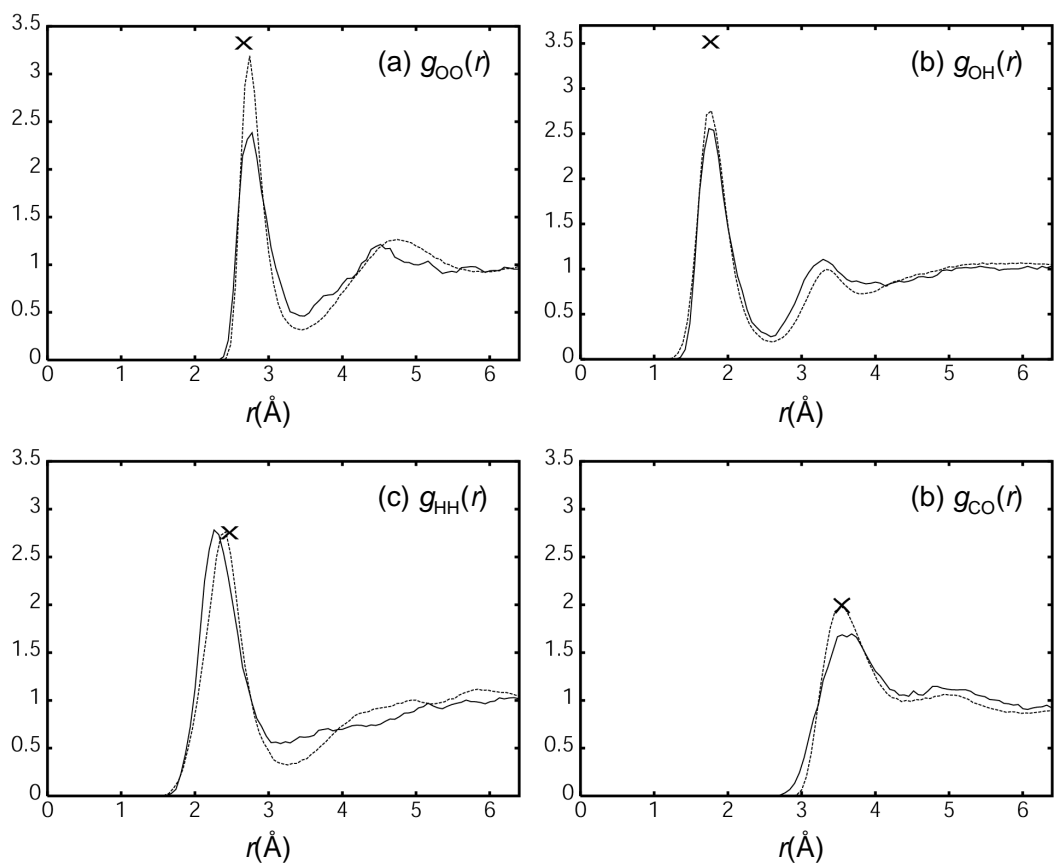}
\caption{
  Calculated hydrogen-bonding and C-O radial distribution functions (solid
  lines).  Dashed line indicate neutron diffraction
  results of Ref.~\cite{YaHi99}. Crosses indicate position of
  the first peak of the RDFs obtained by Haugney et al. using an
  empirical force field\cite{HaFe87}.  }
\label{fig:liq-struct}
\end{figure}

In Fig.~\ref{fig:liq-struct} we have plotted the most characteristic
atom-atom radial distribution functions (RDFs), i.e. the hydrogen
bonding O-O, O-H$_\mathrm{O}$, and H$_{\mathrm{O}}$-H$_\mathrm{O}$ RDF
and the C-O RDF.  For comparison we also plotted results of recent ND
results\cite{YaHi99} and the peak positions obtained using Haugney's
empirical potential\cite{HaFe87}, the latter being considered one of
the most accurate empirical force fields to date.

The pronounced structure in the first three RDFs are a clear
indication of the presence of hydrogen bonds.  Comparison with the
experimental data shows that the positions of the first peaks match
within the statistical error for the O-O and O-H$_\mathrm{O}$ RDFs and
is slightly smaller ($\approx 0.1$~\AA) for the
H$_{\mathrm{O}}$-H$_{\mathrm{O}}$ RDF.  The height of the first peak
is in good agreement for the O-H$_\mathrm{O}$ and
H$_{\mathrm{O}}$-H$_{\mathrm{O}}$ RDFs, that both can be determined
accurately from the ND data.  However, the O-O RDF shows a calculated
first peak height that is significantly lower than the experimental
result. Given the small system of 32 molecules in our simulation, the
discrepancy could well be a system-size effect. On the other hand, the
indirect way by which the the O-O RDF is extracted from ND data could
yield an overestimate of the O-O correlation.  Comparison with force
field results\cite{HaFe87}, that yield significant higher peak values
for the O-O and O-H$_\mathrm{O}$ RDFs, suggests that the Haugney
potential overestimates the hydrogen bonding structure in the liquid,
in line with the observation of Ref.~\cite{YaHi99}.

The number of H-bonds as calculated by integrating the
O-H$_\mathrm{O}$ and O-O RDFs up to the first minimum, and using the
geometrical criterion of Ref.~\cite{HaFe87}, yields values of $1.9$,
$2.0$, and $1.6$, respectively.  This is in good agreement with the
experimental ND results of Ref.~\cite{YaHi99} yielding $1.8$ and
$1.9$ obtained by integrating the O-H$_\mathrm{O}$ and O-O RDFs.
Applying the geometrical criterion to the Haugney force-field
simulation\cite{HaFe87} yields a slightly higher value of $1.9$.

The hydrogen bonding in the liquid phase is accompanied by an
elongation of the OH$_\mathrm{O}$ bond of $0.15$~\AA. A direct comparison
with the experimental results for this change in the geometry of the
methanol molecule is rather difficult due to the large spread in the
reported values.  However, similar change in the geometry is observed
in the DFT-MD study of liquid methanol reported in Ref.~\cite{TsKa99}
and in {\em ab initio} studies of small methanol
clusters\cite{MoYa97}.

The calculated C-O RDF is in reasonable agreement with the ND results,
with the overall shape well reproduced but the first peak clearly less
pronounced than in the ND result. This is consistent with the, in the
previous section found, underestimation of the BLYP binding energy of
the M-O dimer with a C-O distance at the RDF peak position. This is
due to the absence of the dispersion interaction in BLYP.  However, the
absence of the dispersion interaction clearly does not lead to a
completely distorted C-O positional correlation. The comparison of the
calculated C-C RDF with the ND result (not plotted) is very similar.

\subsection*{\normalsize\rm\em 3.2 Dynamics}

\begin{figure}[ht!]
\includegraphics[angle=-90,width=14cm]{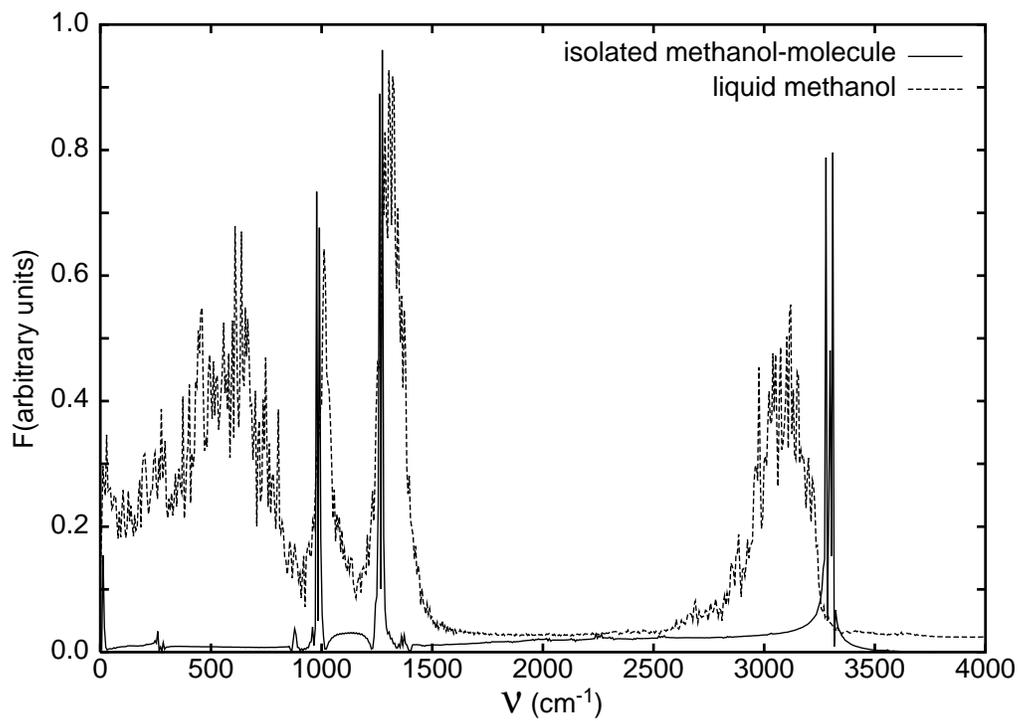}
\caption{Calculated power spectrum of the VACF of the hydroxyl hydrogen for
an isolated methanol at T=200~K (solid line, from Ref.~\cite{ErMe01})
and liquid methanol at T=293~K (dashed line).}
\label{fig:liq-spectrum}
\end{figure}

The time scale of the present simulation (6 ps) allows for an
analysis of the short-time dynamics of liquid methanol.
Figure~\ref{fig:liq-spectrum} shows the power spectrum of the velocity
auto correlation function (VACF) of the hydroxyl hydrogen.  For
comparison we have also plotted the calculated 200~K monomer spectrum of
Ref.~\cite{ErMe01}. The three distinct peaks correspond to the OH
stretch (3100~cm$^{-1}$), C-O-H bend (1600~cm$^{-1}$), and the CO
stretch (1000~cm$^{-1}$). The broad feature below 1000~cm$^{-1}$
indicates the librational-translational (500~cm$^{-1}$) modes of the
methanol molecules.  Compared to the gas phase, the liquid OH stretch
mode has red-shifted by approximately 200 cm$^{-1}$ and broadened
considerably. On the other hand, the C-O-H bending is blue-shifted by
approximately 70 cm$^{-1}$. The observed shifts and broadening are
characteristic for hydrogen bonded liquids and also observed in the
spectrum of water or hydrated methanol.  The calculated shifts compare
reasonably well with experimental infrared spectra\cite{Shim72} that yield
values of $-$354 cm$^{-1}$ and +78 cm$^{-1}$ for the O-H stretch and
C-O-H bend.  The calculated positions and shifts of the modes match
within statistical errors with those of methanol in
aqueous solution\cite{ErMe01} determined using the same computational approach.
This indicates that the intra-molecular dynamics of methanol is
affected in a similar way by an aqueous environment and a methanol
environment.

The diffusion constant $D$ is a key measure of the collective
dynamics.  In view of the limited length of the calculated trajectory
we can only provide a rough estimate. From the mean square
displacement of the oxygen atoms we obtained $D=2.0\pm0.6
\times10^{-9}~m^2/s$, in reasonable agreement with the experimental
value of $2.42\pm0.05\times10^{-9}~m^2/s$ \cite{HuWo80}.

\subsection*{\normalsize\rm\em 3.3  Electronic properties}\label{sec:Results-B}
As the electronic structure is an intrinsic part of a CPMD simulation,
detailed information on the electronic charge distribution is
obtained.  To quantify the charge distribution we used the method of
maximally localized Wannier functions that transforms the Kohn-Sham
orbitals into Wannier functions whose centers (WFC) can be assigned
with a chemical meaning such as being associated with an electron
bonding- or lone-pair (LP)\cite{SiMa98}.

We calculated the positions of the WFCs for the monomer, the dimer,
and 6 independent configurations of the liquid simulation.  
Table~\ref{tab:charges} lists the (average) distances of the
WFCs associated with the oxygen electrons.  Most notably is the small
but significant shift of $0.024$~\AA\ for the OH bond WFC towards the
oxygen atom when going from the monomer to the liquid. At the same
time, one of the LP WFCs shifts away from the oxygen by $0.023~$\AA.
These changes should be considered a manifestation of the hydrogen
bonding and the induced polarization among the dipolar methanol
molecules in the liquid state.

\begin{figure}[ht!]
\includegraphics[angle=-90,width=14.5cm]{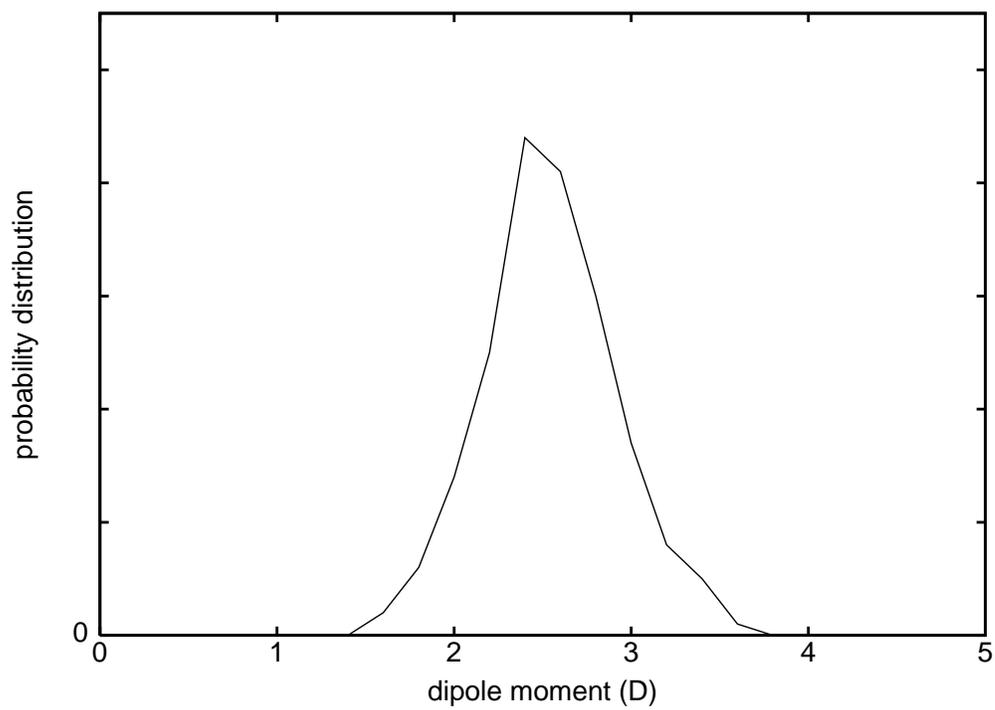}
\caption{Distribution of the molecular dipole moment in liquid methanol,
obtained from 6 independent liquid configurations.}
\label{fig:liq-dipole}
\end{figure}

To quantify the change in the charge distribution in a single number
we calculated the molecular dipole moment assuming the electronic
charge to be distributed as point charges located on the WFCs.  For
liquid water it has been shown that such a partitioning of the charges
over the molecules yields a unique assignment of the WFCs over
distinct molecules\cite{SiPa99-1}. From Table~\ref{tab:dipole}, that
lists the values for the monomer, dimer, and liquid, we observe a
significant enhancement of the dipole moment going from the monomer
via the dimer to the liquid. A comparable liquid-state value of
2.39~D has been observed in a coupled empirical and ab initio MP2
study\cite{TuLa01}.  Note that the value of the dipole moment
is somewhat larger than in the Haugney\cite{HaFe87} (2.33~D) or 
AMBER\cite{AMBER} (2.2~D) force field.
A second important feature of the electronic charge distribution in
the liquid is its fluctuating character due to the thermally
driven configurational changes. In Fig.~\ref{fig:liq-dipole}
we have plotted the calculated distribution of the dipole moments in
the liquid phase. It shows that there is a significant variation
ranging from 1.7~D to 3.5~D.

\section*{Conclusions}
We have demonstrated that \emph{ab initio} MD is a valuable approach
to study the structural, dynamical, and electronic properties of
liquid methanol. The calculated pair distribution functions involving
the hydroxyl hydrogens correlate well with recent state-of-the-art
neutron diffraction experiments of Soper and co-workers. It confirms
their finding that one of the benchmark empirical potentials
overestimates the hydrogen bonding structure.  The calculated
oxygen-oxygen radial distribution function shows significantly less
structure than the experimental neutron diffraction results. Currently
we are studying a larger simulation sample to see whether this is due
to the small system size in the present calculation.  It could also be
a result from an inaccuracy in the experimental result that is
obtained in an indirect way. Results for the dimer binding energies and
the oxygen-carbon RDF suggest that the absence of the dispersion
interaction is notable but has no major impact.  Comparing the
vibrational spectra of the liquid phase against that of the gas phase
monomer shows a significant red shift of the O-H stretch accompanied
by a smaller blue shift of the C-O-H bend mode, in reasonable
agreement with experimental observations with the O-H shift somewhat
underestimated in our calculation. We quantified the electronic charge
distribution using a Wannier function decomposition.  A small but
measurable shift of the positions of the Wannier function centers when
going from the gas-phase to the liquid is accompanied by a
substantial enhancement of the dipole moment. Moreover we have found
that in the liquid the dipole moment fluctuates significantly with
variations up to half the average magnitude. The latter suggest that
the assumption made in empirical potentials using a fixed dipole
moment is a strong simplification.  The present results may be
considered valuable data for improvement of empirical potentials for
the study of liquid methanol.

\begin{ack}
We are grateful to A.K. Soper for providing us data of the RDFs of
Ref.~\cite{YaHi99}.
J.-W.H. and  T.S.v.E acknowledges NWO-CW (Nederlandse Organisatie voor
Wetenschappelijk Onderzoek, Chemische Wetenschappen), J.-W.H through PIONIER.
E.J.M. acknowledges the Royal Netherlands Academy of Art and Sciences for
financial support.  We acknowledge support from the Stichting
Nationale Computerfacileiten (NCF) and the Nederlandse Organisatie
voor Wetenschappelijk Onderzoek (NWO) for the use of supercomputer
facilities. 
\end{ack}


\begin{thebibliography}{10}
\bibitem{YaHi99}
T.~Yamaguchi, K.~Hidaka, A.~K. Soper, Mol. Phys. 96 (1999) 1159; Erratum, 
  Mol. Phys. 97 (1999) 603.

\bibitem{AdBi00}
A.~K. Adya, L.~Bianchi, C.~J. Wormald, J. Chem. Phys. 112 (2000) 4231.

\bibitem{HaFe87}
M.~Haughney, M.~Ferrario, I.~R. McDonald, J. Phys. Chem. 91 (1987) 4934.

\bibitem{BiKa00}
L.~Bianchi, O.~N. Kalugin, A.~K. Adya, C.~J. Wormald, Mol. Simulation 25 (2000)
  321.

\bibitem{TuLa01}
Y.~Tu, A.~Laaksonen, Phys. Rev. E 64 (2001) 026703.

\bibitem{MaSa02}
M.~E. Mart\'{\i}n, M.~L. S\'{a}nchez, F.~J.~O. del Valle, M.~A. Aguilar, 
  J. Chem. Phys. 116 (2002) 1613.

\bibitem{TsKa99}
E.~Tsuchida, Y.~Kanada, M.~Tsukada, Chem. Phys. Lett. 311 (1999) 236.

\bibitem{CaPa85}
R.~Car, M.~Parrinello, Phys. Rev. Lett. 55 (1985) 2471.

\bibitem{LaSp93}
K.~Laasonen, M.~Sprik, M.~Parrinello, R.~Car, J. Chem. Phys. 99 (1993) 9080.

\bibitem{SpHu96}
M.~Sprik, J.~Hutter, M.~Parrinello, J. Chem. Phys. 105 (1996) 1142.

\bibitem{SiPa99-1}
P.~L. Silvestrelli, M.~Parrinello, J. Chem. Phys. 111 (1999) 3572.

\bibitem{MaSp97}
D.~Marx, M.~Sprik, M.~Parrinello, Chem. Phys. Lett. 273 (1997) 360.

\bibitem{ErMe01}
T.~S. van Erp, E.~J. Meijer, Chem. Phys. Lett. 333 (2001) 290.

\bibitem{RaKl02}
S.~Raugei, M.~L. Klein, J. Chem. Phys. 116 (2002) 196.

\bibitem{LeYa88}
C.~Lee, W.~Yang, R.~G. Parr, Phys. Rev. B 37 (1988) 785.

\bibitem{Beck88_2}
A.~D. Becke, Phys. Rev. A 38 (1988) 3098.

\bibitem{MaHu00}
D.~Marx, J.~Hutter, Ab initio molecular dynamics: Theory and implementation,
  in: J.~Grotendorst (Ed.), Modern Methods in Algorithms of Quantum Chemistry,
  Vol.~1 of NIC Series, John von Neumann Insitute for Computing, J\"{u}lich,
  pp. 301-449.

\bibitem{CPMD33}
CPMD, version 3.3, developed by J. Hutter, A. Alavi, T. Deutsch, M. Bernasconi,
  St. Goedecker, D. Marx, M. Tuckerman, and M. Parrinello, MPI f\"ur
  Festk\"orperforschung and IBM Zurich Research Laboratory (1995-1999).

\bibitem{TrMa91}
N.~Troullier, J.~L. Martins, Phys. Rev. B 43 (1991) 1993.

\bibitem{ADF2000}
ADF 2000, G. te Velde, E.~J. Baerends et al. Theoretical Chemistry, Vrije
  Universiteit, Amsterdam.

\bibitem{MoYa97}
O.~M\'{o}, M.~Y\'{a}{\~{n}}ez, J.~Elguero, J. Chem. Phys. 107 (1997) 3592.

\bibitem{SchBr97}
M.~Sch\"{u}tz, S.~Brdarski, P.-O. Widmark, R.~Lindh, G.~Karlstr\"{o}m, J. Chem.
  Phys. 107 (1997) 4597.

\bibitem{GoMo98}
L.~Gonz\'{a}lez, O.~M\'{o}, M.~Y\'{a}{\~{n}}ez, J. Chem. Phys. 109 (1998) 139.

\bibitem{LoBe95}
F.~J. Lovas, S.~P. Belov, M.~Y. Tretyakov, W.~Stahl, R.~D. Suenram, J. Mol.
  Spectrosc. 170 (1995) 478.

\bibitem{LoHa97}
F.~J. Lovas, H.~Hartwig, J. Mol. Spectrosc. 185 (1997) 98.

\bibitem{MoDu99}
W.~T.~M. Mooij, F.~B. van Duijneveldt, J.~G. C.~M. van Duijneveldt-van~de
  Rijdt, B.~P. van Eijck, J. Phys. Chem. A 103 (1999) 9872.

\bibitem{KrPu94}
S.~Kristy\'an, P.~Pulay, Chem. Phys. Lett. 229 (1994) 175.

\bibitem{MeSp96}
E.~J. Meijer, M.~Sprik, J. Chem. Phys. 105 (1996) 8684.

\bibitem{RiBu86}
J.~A. Riddick, W.~B. Bunger, T.~K. Sakano, Organic solvents: physical
  properties and methods of purification, Wiley, New York, 1986.

\bibitem{Nose84_1}
S.~Nos\'e, J. Chem. Phys. 81 (1984) 511.

\bibitem{Shim72}
T.~Shimanouchi, Tables of molecular vibrational frequencies consolidated,
  Volume I, National Bureau of Standards (1972).

\bibitem{HuWo80}
R.~L. Hurle, L.~A. Woolf, Aust. J. Chem. 33 (1980) 1947.

\bibitem{SiMa98}
P.~L. Silvestrelli, N.~Marzari, D.~Vanderbilt, M.~Parrinello, Solid State 
Commun. 107 (1998) 7.

\bibitem{AMBER}
W.~D. Cornell, P.~Cieplak, C.~I. Bayly, I.~R. Gould, K.~M. Merz., Jr., D.~M.
  Ferguson, D.~C. Spellmeyer, T.~Fox, J.~W. Caldwell, P.~A. Kollmann,
  J. Am. Chem. Soc. 117 (1995) 5179.

\bibitem{IvDe53}
E.~V. Ivash, D.~M. Dennison, J. Chem. Phys. 21 (1953) 1804.
\end{thebibliography}

\bibliographystyle{elsart-num}

\clearpage


\begin{table}
\caption{
Complexation energies (kJ/mol) of methanol dimer shown in
Fig.~\ref{fig:dimer}. Numbers are
bare values without zero-point energy corrections and without
entropy contributions.}
\vspace*{10mm}
\begin{tabular}{l|ccc}
CPMD-BLYP  & ADF-BLYP$^a$ & B3LYP$^b$ & MP2$^c$  \\
 \hline
16.4  & 17.3  & 20.6  & 18.4 \\
  \hline
\end{tabular}
\\
{\small
$^a$ Refs.~\cite{ADF2000}\\
$^b$ B3LYP/6-311+G(3df,2p) method. B3LYP/6-311+G(d,p) optimized geometries.
From Ref.~\cite{GoMo98}.\\
$^c$ G2(MP2) method. MP2(full)/6-311+G(d,p) optimized geometries. 
>From Ref.~\cite{GoMo98}.
}
\label{tab:dimer}
\end{table}

\begin{table}
\caption{Electronic charge distribution in 
terms of Wannier function centers. 
\emph{d}(LP) denotes the average distances between
a lone pair WFC and the O nucleus. \emph{d}(OH) and \emph{d}(OC) denote the
(average) distances between the covalent WFC along the O$-$H bond and the 
O$-$C bond with the O nucleus, respectively. 
All distances are given in {\AA}.  Statistical errors for the liquid data 
are around $0.002$.}
\vspace*{10mm}
{\small
\begin{tabular}{lllll}
\hline
\hline
        & \emph{d}(LP) & \emph{d}(LP) & \emph{d}(OH) & \emph{d}(OC)  \\
\hline                                                               
Monomer & 0.305        & 0.305        & 0.533        & 0.562         \\
Dimer   & 0.316        & 0.306        & 0.522        & 0.561         \\
Liquid  & 0.328        & 0.309        & 0.509        & 0.561         \\
\hline
\hline
\\[-10pt]
\end{tabular}
}
\label{tab:charges}
\end{table}

\begin{table}
\caption{Dipole moment.
Experimental value is given in parentheses. Data for the liquid
phase were obtained by averaging over 6 configurations of the MD simulation.
Statistical errors are in the order of some units in the last digit.}
\vspace*{10mm}
{\small
\begin{tabular}{ll}
\hline
\hline
        & $\mu$~(D)         \\ 
\hline
Monomer & 1.73 (1.69$^{a}$) \\
Dimer   & 2.03              \\ 
Liquid  & 2.54              \\
\hline
\hline
\\[-10pt]
$^{a}$Microwave study, ref. \cite{IvDe53}. \\
\end{tabular}
}
\label{tab:dipole}
\end{table}



\end{document}